\documentclass[conference,a4paper]{IEEEtran}
\IEEEoverridecommandlockouts
\usepackage{amsmath}
\usepackage{amssymb}
\usepackage{amsfonts}
\usepackage{slashbox}

\usepackage{graphicx}
\usepackage{cite, amsmath, amsfonts, amssymb, psfrag, epsfig, graphicx}
\newtheorem{definition}{{Definition}}
\newtheorem{theorem}{{Theorem}}

\newtheorem{remark}{{Remark}}

\begin{document}

\title{Adaptation is Useless for Two Discrete Additive-Noise Two-Way Channels}

\author{\IEEEauthorblockN{Lin Song\IEEEauthorrefmark{1},
Fady Alajaji\IEEEauthorrefmark{2},
and Tamas Linder\IEEEauthorrefmark{2}}
\IEEEauthorblockA{\IEEEauthorrefmark{1}\IEEEauthorrefmark{2}Department of Mathematics and Statistics, Queen's University, Ontario, Canada}
\IEEEauthorblockA{Email: \IEEEauthorrefmark{1}lin.song@queensu.ca,\IEEEauthorrefmark{2}\{fady, linder\}@mast.queensu.ca}
\thanks{This work was supported in part by NSERC of Canada.}
}

 \maketitle

\begin{abstract}
In two-way channels, each user transmits and receives at the same time. This allows each encoder to interactively adapt the current input to its own message and all previously received signals. Such coding approach can introduce correlation between inputs of different users, since all the users' outputs are correlated by the nature of the channel. However, for some channels, such adaptation in the coding scheme and its induced correlation among users are useless in the sense that they do not help enlarge the capacity region with respect to the standard coding method (where each user encodes only based on its own message). In this paper, it is shown that adaptation is not helpful for enlarging the capacity region of two classes of two-way discrete channels: the modulo additive-noise channel with memory and the multiple access/degraded broadcast channel.
\end{abstract}


%
\IEEEpeerreviewmaketitle

\section{Introduction}

\begin{figure}[!htp]
\begin{centering}
\includegraphics[width=8.5cm]{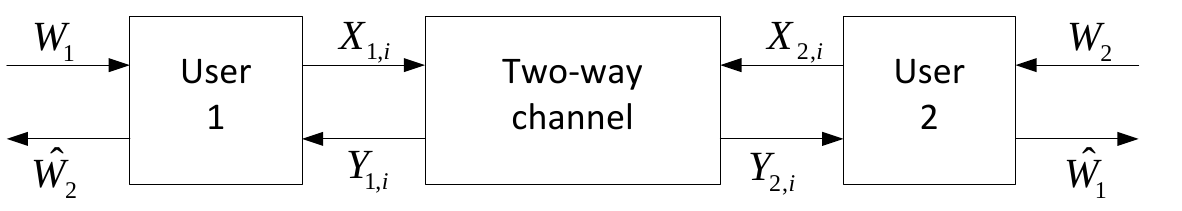}
\caption{Two-way channel.\label{fig:TWModel}}
\end{centering}
\end{figure}

The two-way channel was first introduced by Shannon \cite{Shannon:1961} in 1961. As shown in Fig. \ref{fig:TWModel}, two users (users 1 and 2) exchange messages ($W_1$ and $W_2$) through a common channel. At time~$i$ ($i=1,2,...,n$), the channel input at user $j$ ($j=1,2$) is $X_{j,i}$ and the  output at user $j$ is $Y_{j,i}$. If the two users transmit independent messages, then the first inputs $X_{1,1}$ and $X_{2,1}$ are independent. However, starting from the second use of the channel, user $j$ can adapt the current input $X_{j,i}$ ($i\ge 2$) to all previous outputs $Y_{j,1},...,Y_{j,i-1}$ (this is similar to the case of channels with feedback). This ``adaptation'' coding scheme may introduce dependency between $X_{1,i}$ and $X_{2,i}$ ($i\ge 2$). The dependency between current inputs and previous outputs can be characterized using causal conditioning and the capacity of the two-way channel can be characterized using the directed information \cite{Massey:1990} which is widely used in the characterization of the feedback capacity \cite{Tatikonda:2000, Chen:2005, Permuter:2009, Permuter2:2009, Kim:2008}. However, directed information and causal conditioning result in multi-letter expressions which are not computable. In \cite{Shannon:1961}, Shannon gave single-letter inner and outer bounds on the capacity region of the discrete memoryless two-way channel. Both bounds have the form,
\begin{align*}
R_1 & \le I(X_1;Y_2|X_2),\\
R_2 & \le I(X_2;Y_1|X_1),
\end{align*}
where $R_1$ is the rate of message $W_1$ and $R_2$ is the rate of message $W_2$. For the inner bound $X_1$ and $X_2$ are independent inputs, while for the outer bound $X_1$ and $X_2$ can have arbitrary dependence. Han \cite{Han:1984} gave an improved inner bound by introducing auxiliary random variables and forming a Markov input process. Hekstra and Willems \cite{Hekstra:1989} developed an improved outer bound by introducing an extra dependence balance condition on the set of input distributions. In general, the inner and outer bounds do not coincide. This is because it is difficult to fully characterize the optimal dependency between the inputs of different users in a single-letter expression. 
The single-letter characterization of the capacity region of the discrete memoryless two-way channels is still an open problem.

However, under some circumstances, adaptation is useless from the capacity perspective. Recently, Cheng and Devroye \cite{Devroye:2014} established the capacity regions for some memoryless deterministic two-way channels
(the two-way multiple access/broadcast channel, the two-way Z channel and the two-way interference channel) 
 showing that adapting the current input of each user to all previous outputs will not enlarge the capacity region for these models. The term ``adaptation'' was first coined in \cite{Devroye:2014} for two-way channels.


In this paper, we study noisy versions of two of the deterministic (noiseless) two-way channel models investigated in \cite{Shannon:1961,Devroye:2014}.
Specifically, we consider the  additive-noise two-user two-way channel (2TWC) whose noise processes in both directions are stationary and ergodic, and the additive-noise two-way multiple access/degraded broadcast channel (MA/DBC) whose noise process in the MAC direction is stationary and ergodic while its noise processes in the DBC direction are memoryless. We obtain single-letter capacity region expressions for these channels and show in each case, analogously to \cite{Devroye:2014}, that adapting each user's input to its previously received signals  is useless in terms of enlarging the capacity region over the case of treating the two-way channel as two separate one-way channels concurrently transmitting signals in reverse directions.

This paper is organized as follows. We begin by introducing the channel models in Section \ref{sec:model}. Section \ref{sec:2TWC} considers the discrete additive-noise 2TWC and gives its capacity region, while Section \ref{sec:MADBC} shows the capacity region of the discrete additive-noise MA/DBC. Finally, Section \ref{sec:conclusion} concludes the paper.

\section{Channel Models}\label{sec:model}
In this paper, the discrete (discrete-time finite-alphabet) additive-noise 2TWC channel and the discrete additive-noise MA/DBC are considered. The channel models are introduced in the next two sections.

\subsection{Discrete Additive-Noise 2TWC}\label{sec:model1}
At time $i$, $i=1,...,n$, let $X_{1,i} \in \mathcal{X}_1$, $X_{2,i}\in \mathcal{X}_2$ be the input random variables at user $1$ and $2$, respectively, and let $Y_{1,i}\in \mathcal{Y}_1$ and $Y_{2,i}\in \mathcal{Y}_2$ be the outputs at user $1$ and $2$, respectively. Suppose all the inputs and outputs are from the common finite set, $ \mathcal{X}_1=\mathcal{X}_2=\mathcal{Y}_1=\mathcal{Y}_2=\mathcal{Q}=\{1,2,...,q\}$. The discrete modulo additive-noise 2TWC is described by the following transmission equations:
\begin{align}
Y_{1,i} = X_{1,i}+X_{2,i}+Z_{1,i},\label{eq:2TWC1}\\
Y_{2,i} = X_{1,i}+X_{2,i}+Z_{2,i},\label{eq:2TWC2}
\end{align}
for $i=1,2,...,n$, where ``$+$'' is modulo $q$ addition, $Z_{j,i}\in \mathcal{Q}$, for $j=1,2$. $\{Z_{1,i}\}_{i=1}^n$ and $\{Z_{2,i}\}_{i=1}^n$ are discrete additive-noise processes which are stationary, ergodic, and independent of each other and all the users' messages.

\begin{definition}
An $(n, M_1, M_2)$ code for a discrete additive noise 2TWC consists of two message sets $\mathcal{M}_1=\{1,2,...,M_1\}$, $\mathcal{M}_2=\{1,2,...,M_2\}$, two sequences of encoding functions $f_1^n=(f_{1,1},f_{1,2},...,f_{1,n})$, $f_2^n=(f_{2,1},f_{2,2},...,f_{2,n})$, where $f_{1,1}:  \mathcal{M}_1 \to \mathcal{Q}$, $f_{2,1}:  \mathcal{M}_2 \to \mathcal{Q}$, $f_{1,i}:  \mathcal{M}_1 \times  \mathcal{Q}^{i-1} \to \mathcal{Q}$, $f_{2,i}:  \mathcal{M}_2 \times  \mathcal{Q}^{i-1} \to \mathcal{Q}$ for $i=2,...,n$, and two decoding functions $g_2: \mathcal{M}_2 \times  \mathcal{Q}^{n} \to \mathcal{M}_1$, $g_1: \mathcal{M}_1 \times  \mathcal{Q}^{n} \to \mathcal{M}_2$. 
\end{definition}

Let $W_1 \in \mathcal{M}_1$ and $W_2 \in \mathcal{M}_1$ be the messages transmitted by user 1 and 2, respectively. Suppose $W_1$ and $W_2$ are uniformly distributed on the message sets and are independent of each other. The channel inputs are constructed as follows
\begin{align}
X_{1,1} & =f_{1,1}(W_1),\label{eq:encoder1}\\
X_{2,1} & =f_{2,1}(W_2),\label{eq:encoder2}\\
X_{1,i} & =f_{1,i}(W_1,Y_1^{i-1}), i=2,...,n, \label{eq:encoder3}\\
X_{2,i} & =f_{2,i}(W_2,Y_2^{i-1}), i=2,...,n.\label{eq:encoder4}
\end{align} 
At the end of the $n$th channel use, user 1 and 2 reconstructs messages $W_2$ and $W_1$, respectively, as
\begin{align*}
\hat{W}_2 =g_1(W_1,Y_1^n),\\
\hat{W}_1 = g_2(W_2,Y_2^n).
\end{align*} 
The  error probability of reconstruction at user 1 is defined as $P_{e1}=\text{Pr}\{\hat{W}_2 \neq W_2 \}$ and the  error probability of reconstruction at user 2 is defined as $P_{e2}=\text{Pr}\{\hat{W}_1 \neq W_1 \}$.
\begin{definition}
A rate pair $(R_1,R_2)$ is achievable for a discrete additive-noise 2TWC, if there exists a sequence of $(n, M^{(n)}_1, M^{(n)}_2)$ codes with
\begin{align*}
& \frac{1}{n} \log M^{(n)}_j \ge R_j, j=1,2, n\ge 1,\\
& \lim_{n \to \infty} P^{(n)}_{ej} =0, j=1,2.
\end{align*}
\end{definition}

\begin{definition}
The capacity region of a discrete additive-noise 2TWC is the closure of the convex hull of all achievable rate pairs. 
\end{definition}

\subsection{Discrete Additive-Noise MA/DBC} \label{sub:model2}
\begin{figure}[!htp]
\begin{centering}
\includegraphics[width=8.5cm]{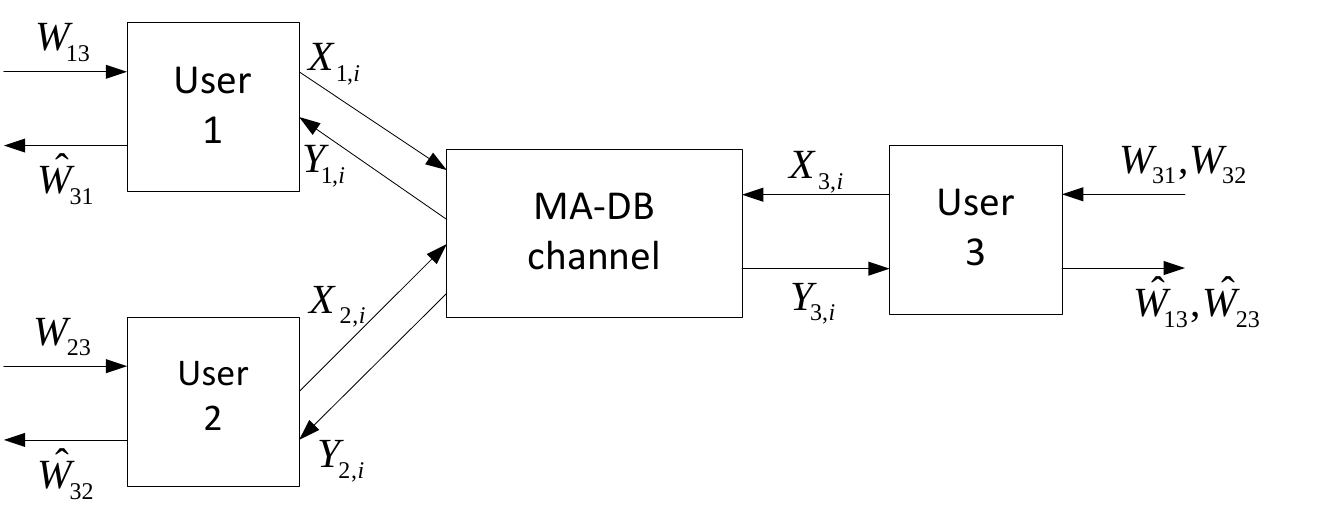}
\caption{Multiple access/degraded broadcast channel.\label{fig:MADBC}}
\end{centering}
\end{figure}

A discrete additive-noise MA/DBC has three users (as shown in Fig. \ref{fig:MADBC}). Users 1 and 2 want to transmit messages $W_{13}$ and $W_{23}$ to user 3. They form a multiple access channel in one direction. User 3 wants to broadcast messages $W_{31}$ and $W_{32}$ to user 1 and 2, respectively, forming a broadcast channel in the other direction. It is also assumed that all inputs and outputs are from the finite set $\mathcal{Q}$. The discrete additive-noise MA/DBC is defined by the following transmission equations: 
\begin{align}
Y_{1,i} &= X_{1,i}+X_{3,i}+Z_{1,i},\label{eq:MADB1}\\
Y_{2,i} &= X_{2,i}+X_{3,i}+Z_{1,i}+Z_{2,i},\label{eq:MADB2}\\
Y_{3,i} &= X_{1,i}+X_{2,i}+X_{3,i}+Z_{3,i},\label{eq:MADB3}
\end{align}
for $i=1,2,...,n$, where ``$+$'' is modulo $q$ addition, $Z_{j,i}\in \mathcal{Q}$, for $j=1,2,3$. $\{Z_{1,i}\}_{i=1}^n$ and $\{Z_{2,i}\}_{i=1}^n$ are discrete memoryless and $\{Z_{3,i}\}_{i=1}^n$ is stationary and ergodic. All the noise processes are mutually independent and all the users' messages.
The definition of channel codes, achievable rates and capacity region are similar to those stated for the discrete additive-noise 2TWC.

\section{Capacity Region of The Discrete Additive-Noise 2TWC}\label{sec:2TWC}
\begin{theorem}
The capacity region of the discrete additive-noise 2TWC as described in Section \ref{sec:model1} is the set of rate pairs $(R_1,R_2)$ such that
\begin{align*}
R_1 \le \log q - \bar{H}(Z_2),\\
R_2 \le \log q - \bar{H}(Z_1).
\end{align*}
where $\bar{H}(\cdot)$ denotes the entropy rate.
\end{theorem}

\subsection{Achievability}\label{sub:TWVach}

Note that $\log q - \bar{H}(Z_2)$ and $\log q - \bar{H}(Z_1)$ are the capacities of the point-to-point discrete one-way additive-noise channels
\begin{align*}
\tilde{Y}_{2,i} & =\tilde{X}_{1,i}+Z_{2,i}, i=1,2,...,n,\\
\tilde{Y}_{1,i} & =\tilde{X}_{2,i}+Z_{1,i}, i=1,2,...,n, 
\end{align*}
respectively. We prove achievability by showing that any coding scheme for the two separate point-to-point discrete additive-noise channels can be used to construct a coding scheme for the discrete additive-noise 2TWC without changing the performance. Therefore, the achievability proof for the discrete additive-noise 2TWC follows directly from the achievability proof for the two separate point-to-point discrete additive-noise channels. Let $\tilde{f}^{(n)}_j: \mathcal{M}^{(n)}_j \to \mathcal{Q}^n$ and $\tilde{g}^{(n)}_j: \mathcal{Q}^n \to \mathcal{M}^{(n)}_j$ be the encoding and decoding functions of the $j$th channel. Then the error probabilities of the two separate point-to-point discrete additive-noise channels are 
$\tilde{P}^{(n)}_{e2}  = \text{Pr}\{\tilde{g}^{(n)}_1(\tilde{f}^{(n)}_1(W_1)+Z^n_2) \neq W_1 \}$ and $\tilde{P}^{(n)}_{e1}  = \text{Pr}\{\tilde{g}^{(n)}_2(\tilde{f}^{(n)}_2(W_2)+Z^n_1) \neq W_2 \}$.
Now we are ready to describe the coding scheme for the discrete additive-noise 2TWC. Let the channel inputs of users 1 and 2 be $X_1^n=\tilde{f}^{(n)}_1(W_1)$ and $X_2^n=\tilde{f}^{(n)}_2(W_2)$, respectively, and reconstruct the messages as 
$\hat{W}_1  = g^{(n)}_2(W_2, Y_2^n) = \tilde{g}^{(n)}_1(Y_2^n-\tilde{f}^{(n)}_2(W_2))$ and $\hat{W}_2  = g^{(n)}_1(W_1, Y_1^n) = \tilde{g}^{(n)}_2(Y_1^n-\tilde{f}^{(n)}_1(W_1))$.
The error probability of reconstruction at user 2 is
\begin{align*}
{P}^{(n)}_{e2} & =  \text{Pr}\{\hat{W}_1 \neq W_1 \}\\
         & =  \text{Pr}\{\tilde{g}^{(n)}_1(Y_2^n-\tilde{f}^{(n)}_2(W_2)) \neq W_1 \}\\
         & =  \text{Pr}\{\tilde{g}^{(n)}_1(X_1^n+X_2^n+Z_2^n-\tilde{f}^{(n)}_2(W_2)) \neq W_1 \}\\
         & =  \text{Pr}\{\tilde{g}^{(n)}_1(\tilde{f}^{(n)}_1(W_1)+Z^n_2) \neq W_1 \}=\tilde{P}^{(n)}_{e2}.
\end{align*}
Similarly, it can be shown that ${P}^{(n)}_{e1}=\tilde{P}^{(n)}_{e1}$. This completes the achievability proof. \hfill \rule{6pt}{6pt}

\subsection{Converse}
For the converse, we start with the achievable rate $R_1$:
\begin{align}
& n R_1  = H(W_1) = H(W_1|W_2) \nonumber\\
    & = I(W_1;Y_2^n|W_2)+H(W_1|W_2,Y_2^n) \nonumber\\
    & \le \sum_{i=1}^n I(W_1;Y_{2,i}|Y_2^{i-1},W_2)+n\epsilon_n \label{eq:2fano}\\
    & = \sum_{i=1}^n H(Y_{2,i}|Y_2^{i-1},W_2)-\sum_{i=1}^n H(Y_{2,i}|Y_2^{i-1},W_{\{1,2\}})+n\epsilon_n\nonumber\\
    & \le n \log q +n\epsilon_n-\sum_{i=1}^n H(Y_{2,i}|Y_2^{i-1},W_{\{1,2\}})\nonumber\\
    & \le n \log q +n\epsilon_n-\sum_{i=1}^n H(Y_{2,i}|Y_2^{i-1},W_{\{1,2\}}, X^i_{\{1,2\}})\nonumber\\
    & =  n \log q +n\epsilon_n-\sum_{i=1}^n H(Z_{2,i}|Z_2^{i-1},W_{\{1,2\}}, X^i_{\{1,2\}})\label{eq:2User1}\\
    & = n \log q +n\epsilon_n- \sum_{i=1}^n H(Z_{2,i}| Z_2^{i-1})\label{eq:2User2}\\
    & = n \log q +n\epsilon_n-H(Z_{2}^n)\nonumber
\end{align}
where $\lim_{n \to \infty}\epsilon_n=0$, $W_{\{1,2\}}$ denotes $(W_1,W_2)$, and $X^i_{\{1,2\}}$ denotes $(X_1^i,X_2^i)$. In the above, \eqref{eq:2fano} follows from Fano's inequality,
 \eqref{eq:2User1} follows from \eqref{eq:2TWC2}, and \eqref{eq:2User2} holds since $Z_{2,i}-Z_2^{i-1}-(W_{\{1,2\}}, X^i_{\{1,2\}})$ form a Markov chain. This Markov chain property holds because the noise process $\{Z_{2,i}\}_{i=1}^n$ is independent of $\{Z_{1,i}\}_{i=1}^n$ and all the users' messages.
Dividing both sides by $n$ and taking the limit yields $R_1 \le \log q - \bar{H}(Z_2)$. It can similarly be shown that $R_2 \le \log q - \bar{H}(Z_1)$.
\hfill \rule{6pt}{6pt}
\begin{remark}
In the capacity-achieving coding scheme, note that $X_{j,i}$ is only a function of $W_j$, $j=1,2$. This fact, together with the converse, implies that ``adaptation'' will not enlarge the capacity region of the discrete additive-noise 2TWC, even when the noise processes have memory. Similar results exist for feedback communication. For example, it is shown in \cite{Alajaji:1995} that feedback does not increase capacity of discrete additive-noise channels with memory (see also \cite{Permuter2:2009,Sen:2011}). 
\end{remark}


\begin{remark}
If the joint noise process $\{(Z_{1,i},Z_{2,i})\}_{i=1}^n$ is memoryless, then the capacity region is the set of rate pairs $(R_1,R_2)$ such that
\begin{align*}
R_1 \le \log q - {H}(Z_2),\\
R_2 \le \log q - {H}(Z_1).
\end{align*}
\end{remark}
This result can be obtained by evaluating Shannon's inner and outer bounds of the 2TWC since the channel is memoryless.

\begin{remark}
A similar result does not hold if the joint noise process $\{(Z_{1,i},Z_{2,i})\}_{i=1}^n$ has memory. For example, if $\mathcal{Q}$ is binary, $\{Z_{1,i}\}_{i=1}^n$ is memoryless and uniformly distributed, and $Z_{2,i}=Z_{1,i-1}$, then $(\log q - \bar{H}(Z_2),\log q - \bar{H}(Z_1))=(0,0)$. However, the rate pair $(1,0)$ can be achieved by letting $X_{1,0}=X_{2,0}=0$, $X_{1,i}=U_i+X_{1,i-1}+Y_{1,i-1}$, $X_{2,i}=0$ for $i=1,...,n$, where $Y_{1,0}=0$, $\{U_i\}_{i=1}^n$ is the binary information sequence which is transmitted from user 1 to user 2. This is because under the proposed coding scheme, the received signal at user 2 is 
\begin{align*}
& Y_{2,i}  = X_{1,i} + X_{2,i} + Z_{2,i} \\
        & = U_i+X_{1,i-1}+Y_{1,i-1} + Z_{2,i}\\
        & = U_i+X_{1,i-1}+X_{1,i-1} + Z_{1,i-1} + Z_{2,i} = U_i.
\end{align*}
\end{remark}

\section{Capacity Region of The Discrete Additive-Noise MA/DBC}\label{sec:MADBC}
In this section, we show that ``adaptation'' is useless for the discrete additive-noise MA/DBC. The following theorem characterizes the capacity region of this channel.
\begin{theorem}
The capacity region of sending uniform and independent messages $(W_{13},W_{23},W_{31},W_{32})$ over the discrete memoryless additive-noise MA/DBC as described in Section~\ref{sub:model2} is the closure of the convex hull of all rate quadruples $(R_{13},R_{23},R_{31},R_{32})$ satisfying
\begin{align}
R_{13} + R_{23} & \le \log q - \bar{H}(Z_3)\label{eq:Mcapacity1}\\
R_{31} & \le I(X_3;X_3+Z_1|U)\label{eq:Mcapacity2}\\
R_{32} & \le I(U;X_3+Z_1+Z_2)\label{eq:Mcapacity3}
\end{align}
for some $p(u,x_3)$, where $\bar{H}(\cdot)$ denotes the entropy rate, and $| \mathcal{U}| \le q+1$.
\end{theorem}

\subsection{Achievability}
The proof of achievability of the MA/DBC is similar to that of the 2TWC in Section. \ref{sub:TWVach}. Note that \eqref{eq:Mcapacity1} describes the capacity region of the discrete additive-noise multiple access channel (MAC) \cite{Ahlsweda:1971, Liao:1972},
$$\tilde{Y}_{3,i} = \tilde{X}_{1,i}+ \tilde{X}_{2,i}+Z_{3,i}, i=1,2,...,n.$$
Furthermore, \eqref{eq:Mcapacity2} and \eqref{eq:Mcapacity3} give the capacity region of  the discrete additive-noise degraded broadcast channel (DBC) \cite{Bergmans:1973,Gallager:1974}
\begin{align*}
\tilde{Y}_{1,i} &= \tilde{X}_{3,i}+Z_{1,i},\\
\tilde{Y}_{2,i} &= \tilde{X}_{3,i}+Z_{1,i}+Z_{2,i}.
\end{align*}
We prove the achievablility by showing that separate coding schemes for the MAC and DBC can be used to construct a coding scheme for the MA/DBC with identical performance. Let $\tilde{f}^{(n)}_{1}: \mathcal{M}^{(n)}_{13} \to \mathcal{Q}^n$, $\tilde{f}^{(n)}_{2}: \mathcal{M}^{(n)}_{23} \to \mathcal{Q}^n$ and $\tilde{g}^{(n)}_{3}: \mathcal{Q}^n \to \mathcal{M}^{(n)}_{13} \times \mathcal{M}^{(n)}_{23}$ be the encoding and decoding functions of the MAC. Let $\tilde{f}^{(n)}_{3}: \mathcal{M}^{(n)}_{31}\times \mathcal{M}^{(n)}_{32} \to \mathcal{Q}^n$, $\tilde{g}^{(n)}_{1}: \mathcal{Q}^n \to \mathcal{M}^{(n)}_{31} $ and $\tilde{g}^{(n)}_{2}: \mathcal{Q}^n \to \mathcal{M}^{(n)}_{32}$ be the encoding and decoding functions of the DBC. 
The error probabilities of these two one-way channels are 
\begin{align*}
&\tilde{P}^{(n)}_{e3} \\
& = \text{Pr}\{\tilde{g}^{(n)}_3(\tilde{f}^{(n)}_1(W_{13})+\tilde{f}^{(n)}_2(W_{23})+Z^n_3) \neq (W_{13},W_{23}) \}, \\
&\tilde{P}^{(n)}_{e1} = \text{Pr}\{\tilde{g}^{(n)}_1(\tilde{f}^{(n)}_3(W_{31},W_{32})+Z^n_1) \neq W_{31} \},\\
&\tilde{P}^{(n)}_{e2} = \text{Pr}\{\tilde{g}^{(n)}_2(\tilde{f}^{(n)}_3(W_{31},W_{32})+Z^n_1+Z^n_2) \neq W_{32} \}.
\end{align*}
We construct the coding scheme for the MA/DBC based on $\tilde{f}^{(n)}_j$ and $\tilde{g}^{(n)}_j$, $j=1,2,3$.  Let the channel inputs be $X_1^n=\tilde{f}^{(n)}_1(W_{13})$,  $X_2^n=\tilde{f}^{(n)}_2(W_{23})$ and $X_3^n=\tilde{f}^{(n)}_3(W_{13},W_{23})$, and define the reconstructions as
\begin{align*}
&(\hat{W}_{13},\hat{W}_{23}) = g^{(n)}_3(W_{31},W_{32}, Y_3^n) \\
&\qquad \quad \qquad= \tilde{g}^{(n)}_3(Y_3^n-\tilde{f}^{(n)}_3(W_{31},W_{32})),\\
&\hat{W}_{31} = g^{(n)}_1(W_{13}, Y_1^n) = \tilde{g}^{(n)}_1(Y_1^n-\tilde{f}^{(n)}_1(W_{13})),\\
&\hat{W}_{32}  = g^{(n)}_2(W_{23}, Y_2^n) = \tilde{g}^{(n)}_2(Y_2^n-\tilde{f}^{(n)}_2(W_{23})).
\end{align*}
The error probability of reconstruction at user 3 is 
\begin{align*}
&{P}^{(n)}_{e3} =  \text{Pr}\{(\hat{W}_{13},\hat{W}_{23}) \neq (W_{13},W_{23})\}\\
         & =  \text{Pr}\{\tilde{g}^{(n)}_3(Y_3^n-\tilde{f}^{(n)}_3(W_{31},W_{32})) \neq (W_{13},W_{23}) \}\\
         & =  \text{Pr}\{\tilde{g}^{(n)}_3(X_1^n+X_2^n+X_3^n+Z_3^n-X_3^n) \neq (W_{13},W_{23})\}\\
         & =  \text{Pr}\{\tilde{g}^{(n)}_3(\tilde{f}^{(n)}_1(W_{13})+\tilde{f}^{(n)}_2(W_{23})+Z^n_3) \neq (W_{13},W_{23}) \}\\
         & =  \tilde{P}^{(n)}_{e3}.
\end{align*}
Similarly, it can be shown ${P}^{(n)}_{e1}=\tilde{P}^{(n)}_{e1}$ and  ${P}^{(n)}_{e2}=\tilde{P}^{(n)}_{e2}$, which completes the proof of the achievability.
\hfill \rule{6pt}{6pt}
\subsection{Converse}
For the achievable rates $R_{13}$ and $R_{23}$, we have
\begin{align}
&  nR_{13} + n R_{23} = H(W_{13},W_{23})= H(W_{13},W_{23}| W_{31},W_{32})\nonumber\\
         & = I(W_{13},W_{23};Y_3^n| W_{\{31,32\}}) + H(W_{13},W_{23}|Y_3^n, W_{\{31,32\}})\nonumber\\
         & \le I(W_{13},W_{23};Y_3^n| W_{\{31,32\}}) + n \epsilon_{n} \label{eq:fano2}\\
         & = \sum_{i=1}^n I(W_{13},W_{23};Y_{3,i}|Y_3^{i-1},W_{\{31,32\}}) + n\epsilon_{n} \nonumber\\
         & = \sum_{i=1}^n  H(Y_{3,i}|Y_3^{i-1},W_{\{31,32\}})\nonumber\\
         & \quad -\sum_{i=1}^n H(Y_{3,i}|Y_3^{i-1},W_{\{13,23,31,32\}})  + n\epsilon_{n} \nonumber\\
         & \le n \log q+ n \epsilon_{n}  - \sum_{i=1}^n H(Y_{3,i}|Y_3^{i-1},W_{\{13,23,31,32\}},X^i_{\{1,2,3\}}) \nonumber \\
         & = n \log q+ n \epsilon_{n}  - \sum_{i=1}^n H(Z_{3,i}|Z_3^{i-1},W_{\{13,23,31,32\}},X^i_{\{1,2,3\}}) \label{eq:3User1} \\
         & = n\log q+ n \epsilon_{n} - \sum_{i=1}^n H(Z_{3,i}|Z_{3}^{i-1}) \label{eq:3User2} \\
         & = n\log q-  H(Z_{3}^n) +n \epsilon_{n} \nonumber
\end{align}
where $\lim_{n \to \infty}\epsilon_{n}=0$. Here \eqref{eq:fano2} follows from Fano's inequality, \eqref{eq:3User1} follows from \eqref{eq:MADB3}, and \eqref{eq:3User2} holds since $Z_{3,i}-Z_{3}^{i-1}-(W_{\{13,23,31,32\}},X^i_{\{1,2,3\}})$ form a Markov chain which holds because
 $\{Z_{3,i}\}_{i=1}^n$ is independent of  $\{Z_{1,i}\}_{i=1}^n$, $\{Z_{2,i}\}_{i=1}^n$ and all the users' messages.
Dividing both sides by $n$ and taking the limit yields $R_{13} + R_{23} \le \log q - \bar{H}(Z_3)$.

For achievable rate $R_{32}$, we write
\begin{align}
&  nR_{32}  = H(W_{32}) = H(W_{32}|W_{23})\nonumber\\
        & = I(W_{32};Y_2^n|W_{23}) + H(W_{32}|Y_2^n,W_{23}) \nonumber\\
        & \le I(W_{32};Y_2^n|W_{23}) + n \hat{\epsilon}_{n} \nonumber \\
        & = \sum_{i=1}^n I(W_{32};Y_{2,i}|Y_2^{i-1},W_{23}) + n \hat{\epsilon}_{n} \nonumber\\
        & = \sum_{i=1}^n I(W_{32};Y_{2,i}|Y_2^{i-1},W_{23},X_2^{i}) + n \hat{\epsilon}_{n}\nonumber\\
        & = \sum_{i=1}^n I(W_{32};\tilde{Y}_{2,i}|\tilde{Y}_2^{i-1},W_{23},X_2^{i}) + n \hat{\epsilon}_{n} \nonumber 
        \\
        & = \sum_{i=1}^n I(W_{32};\tilde{Y}_{2,i}|\tilde{Y}_2^{i-1},W_{23}) + n \hat{\epsilon}_{n}\label{eq:2R3}\\
        & \le \sum_{i=1}^n I(W_{32},\tilde{Y}_2^{i-1},W_{23};\tilde{Y}_{2,i}) +n \hat{\epsilon}_{n} \nonumber\\
        & \le \sum_{i=1}^n I(W_{32},W_{23},W_{13},\tilde{Y}_2^{i-1},\tilde{Y}_1^{i-1};\tilde{Y}_{2,i}) + n \hat{\epsilon}_{n} \label{eq:2R5}
\end{align}
where $\lim_{n \to \infty}\hat{\epsilon}_{n}=0$,  $\tilde{Y}_{2,i}$ denotes $X_{3,i}+Z_{1,i}+Z_{2,i}$, 
$\tilde{Y}^{i-1}_{2}$ denotes $X_3^{i-1}+Z_1^{i-1}+Z_2^{i-1}$ and $\tilde{Y}_1^{i-1}$ denotes $X_3^{i-1}+Z_1^{i-1}$. \eqref{eq:2R3} holds since $X_2^{i}$ is a function of $(\tilde{Y}_2^{i-1},W_{23})$.

Now, we consider the achievable rate $R_{31}$:
\begin{align}
&  nR_{31}  = H(W_{31}) = H(W_{31}|W_{32},W_{13},W_{23}) \nonumber \\
        & = I(W_{31};Y_1^n|W_{\{32,13,23\}}) + H(W_{31}|W_{\{32,13,23\}},Y_1^n) \nonumber \\
        & \le I(W_{31};Y_1^n,Y_2^n|W_{\{32,13,23\}}) + n \tilde{\epsilon}_{n} \nonumber\\
        & = \sum_{i=1}^n I(W_{31};Y_{1,i},Y_{2,i}|Y_1^{i-1},Y_2^{i-1},W_{\{32,13,23\}}) + n \tilde{\epsilon}_{n}\nonumber\\
        & \le \sum_{i=1}^n I(W_{31},X_{3,i};Y_{\{1,2\},i}|Y_{\{1,2\}}^{i-1},W_{\{32,13,23\}}) + n \tilde{\epsilon}_{n}\nonumber\\
        & = \sum_{i=1}^n I(W_{31},X_{3,i};Y_{\{1,2\},i}|Y_{\{1,2\}}^{i-1},W_{\{32,13,23\}},X_{\{1,2\}}^i) \nonumber\\
        &\quad + n\tilde{\epsilon}_{n}\nonumber\\
        & = \sum_{i=1}^n I(W_{31},X_{3,i};\tilde{Y}_{\{1,2\},i}|\tilde{Y}_{\{1,2\}}^{i-1},W_{\{32,13,23\}},X_{\{1,2\}}^i)\nonumber\\
        & \quad + n\tilde{\epsilon}_{n}\nonumber\\
        & = \sum_{i=1}^n I(W_{31},X_{3,i};\tilde{Y}_{\{1,2\},i}|\tilde{Y}_{\{1,2\}}^{i-1},W_{\{32,13,23\}}) + n\tilde{\epsilon}_{n}\nonumber\\
        & = \sum_{i=1}^n I(X_{3,i};\tilde{Y}_{\{1,2\},i}|\tilde{Y}_{\{1,2\}}^{i-1},W_{\{32,13,23\}})  \nonumber\\
        & \quad + \sum_{i=1}^n I(W_{31};\tilde{Y}_{\{1,2\},i}|\tilde{Y}_{\{1,2\}}^{i-1},W_{\{32,13,23\}},X_{3,i}) + n\tilde{\epsilon}_{n}\nonumber\\
        & = \sum_{i=1}^n I(X_{3,i};\tilde{Y}_{\{1,2\},i}|\tilde{Y}_{\{1,2\}}^{i-1},W_{\{32,13,23\}})  +n\tilde{\epsilon}_{n} \label{eq:3R1}\\
        & = \sum_{i=1}^n I(X_{3,i};\tilde{Y}_{1,i}|\tilde{Y}_{\{1,2\}}^{i-1},W_{\{32,13,23\}}) + n\tilde{\epsilon}_{n} \label{eq:3R2}
\end{align}
where $\lim_{n \to \infty}\tilde{\epsilon}_{n}=0$. Here \eqref{eq:3R1} holds because  $W_{31}-(\tilde{Y}_{\{1,2\}}^{i-1},W_{\{32,13,23\}},X_{3,i})-\tilde{Y}_{\{1,2\},i}$ form a Markov chain, and 
\eqref{eq:3R2} holds since
$\tilde{Y}_{2,i}-(\tilde{Y}_{1,i},\tilde{Y}_{\{1,2\}}^{i-1},W_{\{32,13,23\}})-X_{3,i}$ form a Markov chain.
These Markov chain properties hold since $\{Z_{1,i}\}_{i=1}^n$ and $\{Z_{2,i}\}_{i=1}^n$ are discrete memoryless and independent of each other and all the users' messages,  $\tilde{Y}_{1,i}=X_{3,i}+Z_{1,i}$, and $\tilde{Y}_{2,i}=\tilde{Y}_{1,i}+Z_{2,i}$.
Let $U_i=(\tilde{Y}_{\{1,2\}}^{i-1},W_{\{32,13,23\}})$, so that $U_i-X_{3,i}-\tilde{Y}_{\{1,2\},i}$ form a Markov chain and from \eqref{eq:2R5} and \eqref{eq:3R2} we have 
$n R_{32}  \le \sum_{i=1}^n I(U_i;\tilde{Y}_{2,i}) + n \hat{\epsilon}_{n}$ and $nR_{31}  \le \sum_{i=1}^n I(X_{3,i};\tilde{Y}_{1,i}|U_i) + n\tilde{\epsilon}_{n}$.
Define the time-sharing random variable $K$ to be uniform over $\{1,2,...,n\}$ and independent of all messages, inputs and outputs, and let $U=(K,U_K)$, $X_3=X_{3,K}$, $Z_1=Z_{1,K}$, $Z_2=Z_{2,K}$ $\tilde{Y}_1=X_3+Z_1=\tilde{Y}_{1,K}$, $\tilde{Y}_2=X_3+Z_1+Z_2=\tilde{Y}_{2,K}$. Then we have
\begin{align*}
n R_{32} & \le \sum_{i=1}^n I(U_i;\tilde{Y}_{2,i}) + n \hat{\epsilon}_{n} = nI(U_K;\tilde{Y}_{2,K}|K)+ n \hat{\epsilon}_{n}\\
& \le nI(U;\tilde{Y}_2) +\hat{\epsilon}_{n}=nI(U;X_3+Z_1+Z_2) +n\hat{\epsilon}_{n},\\
nR_{31} & \le \sum_{i=1}^n I(X_{3,i};\tilde{Y}_{1,i}|U_i) + n\tilde{\epsilon}_{n}=n I(X_{3};\tilde{Y}_{1}|U)+ n\tilde{\epsilon}_{n}\\
& = n I(X_{3};X_3+Z_1|U)+ n\tilde{\epsilon}_{n}.
\end{align*}
The bound on the cardinality of $U$ can be established using the convex cover method \cite{Gamal:2012}.
\hfill \rule{6pt}{6pt}
\begin{remark}
In the achievability coding scheme, inputs at each user are functions of its own message(s). This fact together with the converse imply that ``adaptation'' will not enlarge the capacity region of the discrete additive-noise MA/DBC. 
\end{remark}

\section{Conclusion}\label{sec:conclusion}
In this work, we have shown that for the discrete additive-noise 2TWC and the MA/DBC, adaptation is useless from a capacity region perspective. For these two classes of channels, standard coding schemes where each user encodes based on its own message(s) can achieve optimality and the capacity regions can be characterized by single-letter expressions. However, even for these channels, adaptation can be beneficial in terms of reducing the complexity of coding schemes and improving error exponents. Identifying conditions under which adaptation is useless for a general two-way network is still an open problem. 

\end{document}